\documentclass[useAMS,usenatbib]{mn2e}

\usepackage{verbatim} 
\usepackage{natbib} 
\usepackage{amsmath} 
\usepackage{amsbsy}
\usepackage{amssymb}
\usepackage{mathrsfs} 
\usepackage{lscape} 
\usepackage{graphicx}
\usepackage{epstopdf}
\usepackage{deluxetable}
\usepackage{ctable} 
\usepackage{fixltx2e}

\title[Neutral Hydrogen in Quasar-Mass Halos]{A Stellar Feedback Origin for Neutral Hydrogen in High-Redshift Quasar-Mass Halos}
\author[Faucher-Gigu\`ere et al.]{Claude-Andr\'e Faucher-Gigu\`ere,$^{1}$\thanks{cgiguere@northwestern.edu} Robert Feldmann,$^{2}$ Eliot Quataert,$^{2}$ Du\v{s}an\newauthor Kere\v{s},$^{3}$ Philip F. Hopkins$^{4}$ and Norman Murray$^{5,6}$\vspace*{6pt}\\
$^{1}$Department of Physics and Astronomy and Center for Interdisciplinary Exploration and Research in Astrophysics (CIERA),\\ Northwestern University, 2145 Sheridan Road, Evanston, IL 60208, USA: cgiguere@northwestern.edu.\\
$^{2}$Department of Astronomy and Theoretical Astrophysics Center, University of California, Berkeley, CA 94720-3411, USA.\\
$^{3}$Department of Physics, Center for Astrophysics and Space Sciences, University of California, San Diego, 9500 Gilman Drive,\\ La Jolla, CA 92093.\\
$^{4}$TAPIR, Mailcode 350-17, California Institute of Technology, Pasadena, CA 91125, USA.\\
$^{5}$Canadian Institute for Theoretical Astrophysics, 60 St. George Street, University of Toronto, ON M5S 3H8, Canada.\\
$^{6}$Canada Research Chair in Astrophysics.
}

\begin{document}
\maketitle


\begin{abstract}
Observations reveal that quasar host halos at $z\sim2$ have large covering fractions of cool dense gas ($\gtrsim60$\% for Lyman limit systems within a projected virial radius). Most simulations have so far have failed to explain these large observed covering fractions. We analyze a new set of 15 simulated massive halos with explicit stellar feedback from the FIRE project, covering the halo mass range $M_{\rm h}\approx 2\times10^{12}-10^{13}$ M$_{\odot}$ at $z=2$. This extends our previous analysis of the circum-galactic medium of high-redshift galaxies to more massive halos. AGN feedback is not included in these simulations. We find Lyman limit system covering fractions consistent with those observed around quasars. The large HI covering fractions arise from star formation-driven galactic winds, including winds from low-mass satellite galaxies that interact with cosmological filaments. We show that it is necessary to resolve these satellite galaxies and their winds to reproduce the large Lyman limit system covering fractions observed in quasar-mass halos. Our simulations predict that galaxies occupying dark matter halos of mass similar to quasars but without a luminous AGN should have Lyman limit system covering fractions comparable to quasars. 
\end{abstract}

\begin{keywords}
galaxies: formation --- galaxies: evolution --- galaxies: haloes --- quasars: absorption lines --- intergalactic medium --- cosmology: theory\vspace{-0.5cm}
\end{keywords}

\section{Introduction}
Spectroscopic measurements of gas flows around galaxies using sight lines to background quasars 
provide one of the most direct ways of probing the cosmological inflows and galactic outflows that regulate galaxy growth. 
Over the past several years, this technique has been used at both low redshift and around the peak of the cosmic star formation history at $z\gtrsim2$ \citep[e.g.,][]{2003ApJ...584...45A, 2006ApJ...651...61H, 2010ApJ...717..289S, 2011Sci...334..948T, 2014MNRAS.445..794T}. 
The technique has also been applied to a wide range of foreground objects, including dwarf galaxies \citep[e.g.,][]{2014ApJ...796..136B}, damped Ly$\alpha$ absorbers \citep[e.g.,][]{2015ApJ...808...38R}, luminous red galaxies \citep[LRGs; e.g.,][]{2010ApJ...716.1263G}, $\sim L^{\star}$ star-forming galaxies \citep[e.g.,][]{2012ApJ...750...67R}, and quasars \citep[e.g.,][]{2013ApJ...762L..19P}. 
Driven by this explosion in high-quality observations, many groups have used cosmological simulations to make predictions for circum-galactic medium (CGM) absorbers \citep[e.g.,][]{2011MNRAS.412L.118F, 2011MNRAS.413L..51K, 2011MNRAS.418.1796F, 2012MNRAS.424.2292G, 2012MNRAS.425.1270S, 2013ApJ...765...89S, 2013MNRAS.430.1548H, 2015MNRAS.448..895S}. 
Such comparisons are particularly valuable as state-of-the-art cosmological galaxy formation models have now broadly converged on their predictions for the global stellar properties of galaxy populations but diverge strongly on their predictions for gas properties \citep[][]{2015ARA&A..53...51S}. 
Thus, CGM observations can break degeneracies between galaxy formation theories.

Our focus in this Letter is on the CGM of galaxies likely to be traced by luminous quasars at $z\sim2$, which have a characteristic halo mass $M_{\rm h}\sim10^{12.5}$ M$_{\odot}$ \citep[e.g.,][]{2012MNRAS.424..933W}. 
\citet[][hereafter PHS13]{2013ApJ...762L..19P} reported a surprisingly high covering fraction $f_{\rm cov}(>10^{17.2};~<R_{\rm vir})\approx0.64^{+0.06}_{-0.07}$ of Lyman limit systems (LLSs; $N_{\rm HI}>10^{17.2}$ cm$^{-2}$) within a projected virial radius of  $z\sim2-2.5$ quasars \citep[see also][]{2014ApJ...796..140P}. 
The high covering fraction of cool gas in quasar halos is in contrast to the lower fraction $f_{\rm cov}(10^{17.2};~<R_{\rm vir})=0.30\pm0.14$ measured by \cite{2012ApJ...750...67R} around $z\sim2-2.5$ Lyman break galaxies (LBGs) in the Keck Baryonic Structure Survey (KBSS). 
The LBGs in KBSS reside in dark matter halos of characteristic mass $M_{\rm h}\approx10^{12}$ M$_{\odot}$ \citep[][]{2005ApJ...620L..75A}, a factor $\sim3$ lower than luminous quasars. 
Using cosmological zoom-in simulations of galaxy formation with stellar feedback but neglecting the effects of active galactic nuclei (AGN), \citet[][]{2014ApJ...780...74F} and \citet[][hereafter FG15]{2015MNRAS.449..987F} showed that the LLS covering fractions in the simulations were broadly consistent with those measured in LBG halos (see also Shen et al. 2013\nocite{2013ApJ...765...89S}). 
Both studies however concluded that the most massive halos in their analyses could not explain the LLS covering fractions measured around quasars (but see Rahmati et al. 2015\nocite{2015MNRAS.446..521S}, which we discuss further in \S \ref{sec:lls_properties}). 

In this work, we extend the FG15 analysis with a new set of 15 halos simulated to $z=2$ with stellar feedback physics from the FIRE (``Feedback In Realistic Environments'') project and with masses representative of quasar hosts.\footnote{See project website: http://fire.northwestern.edu} 
These simulations are part of the MassiveFIRE simulation suite described in Feldmann et al. (2016)\nocite{2016MNRAS.458L..14F}. 
We use these simulations to revisit the comparison with HI covering fractions measured around $z\sim2$ quasars. 
The simulations we analyze here do not include AGN but comparing them observations of quasar-mass halos is useful because it can show whether the observed covering fractions in quasar halos require the presence of a luminous AGN or not. 

We describe our simulations and analysis methodology in \S \ref{sec:sims}, discuss our main results in \S \ref{sec:results}, and conclude in \S \ref{sec:conclusion}. Throughout, we assume a standard $\Lambda$CDM cosmology with parameters consistent with the latest constraints ($h\approx0.7$, $\Omega_{\rm m} = 1-\Omega_{\Lambda} \approx 0.27$ and $\Omega_{\rm b}\approx0.046$; Planck 2015)\nocite{Planck2015}.

\begin{figure*}
\begin{center}
\includegraphics[width=0.8\textwidth]{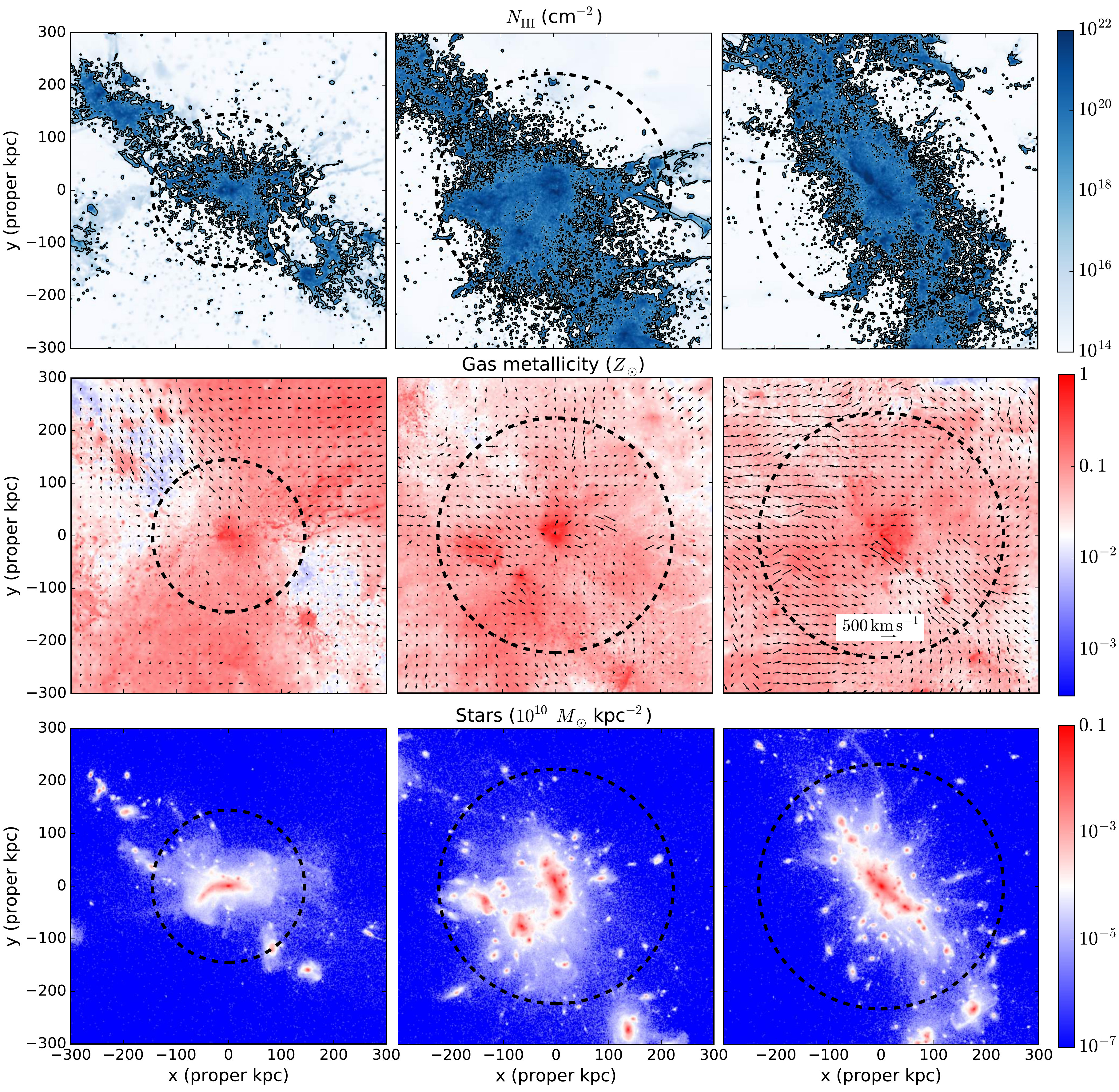}
\end{center}
\caption[]{HI column density (top), gas-phase metallicity (middle) and stellar mass surface density (bottom) maps for three representative MassiveFIRE halos at $z=2$ (from left to right: $M_{\rm h}(z=2)=(2.4, 8.8, 9.9)\times10^{12}$ M$_{\odot}$). 
The virial radius is indicated by dashed circle in each panel and solid contours indicate Lyman limit systems. 
The vectors on metallicity maps indicate projected mass-weighted velocities.
}
\label{fig:halo_images} 
\end{figure*}

\begin{figure}
\begin{center}
\includegraphics[width=0.47\textwidth]{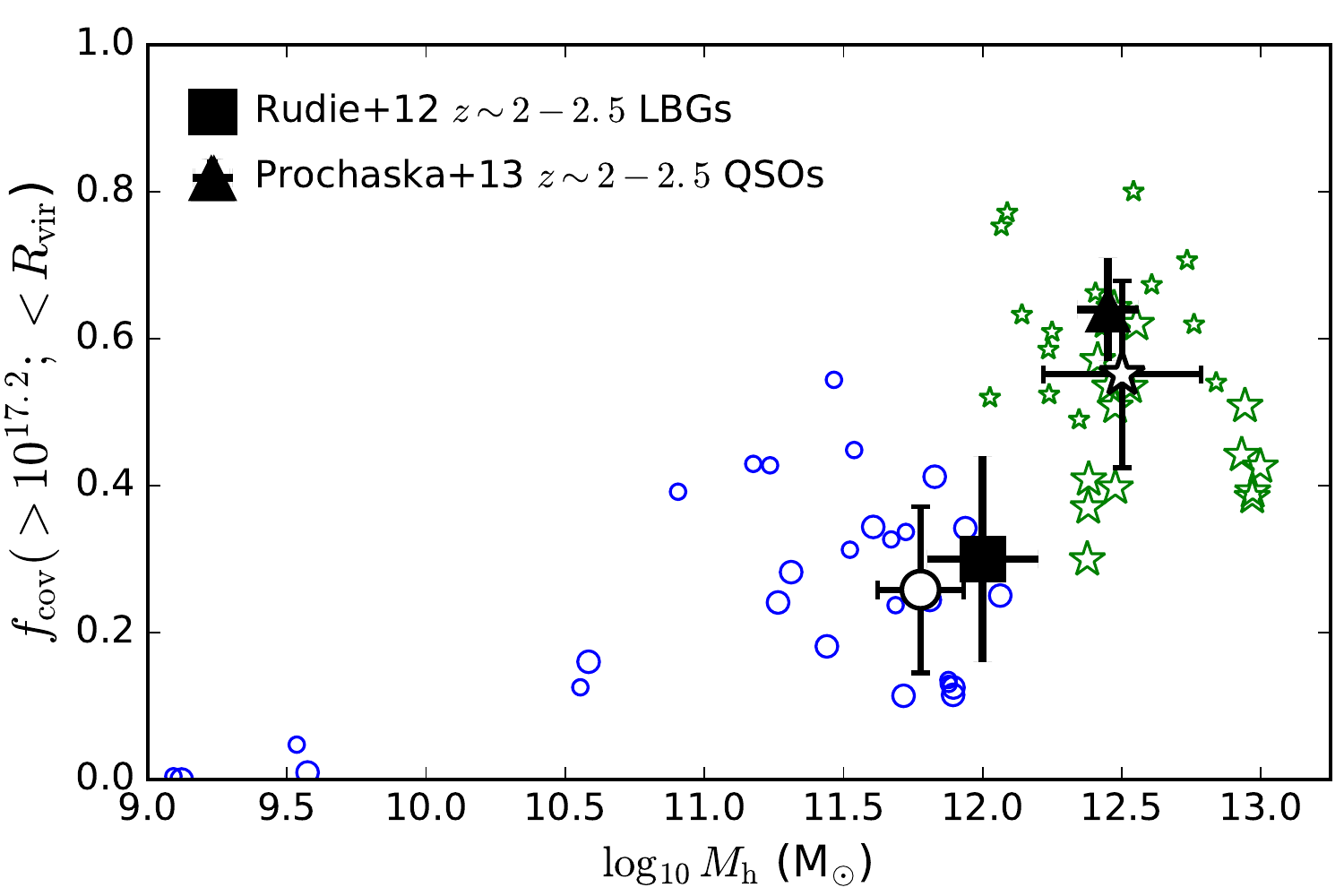}
\end{center}
\caption[]{\emph{Blue circles}: Lyman limit system covering fractions at $z=2$ (large) and $z=2.5$ (small) within a projected virial radius for the simulated halos analyzed in FG15 (HR resolution or better only). 
\emph{Green stars:} Covering fractions at $z=2$ (large) and $z=2.5$ (small) for the MassiveFIRE halos. 
The open black symbols show averages over simulated LBG-mass halos and quasar-mass halos with the error bars showing the standard deviations of the simulated data points included in the averages.
We compare the simulated covering fractions to LLS measurements transverse to LBGs at $z\sim2-2.5$ by \cite{2012ApJ...750...67R} (black square) and transverse to  luminous quasars in the same redshift interval by \cite{2013ApJ...762L..19P} (black triangle). 
}
\label{fig:fcov_summary} 
\end{figure}

\begin{figure*}
\begin{center}
\includegraphics[width=0.8\textwidth]{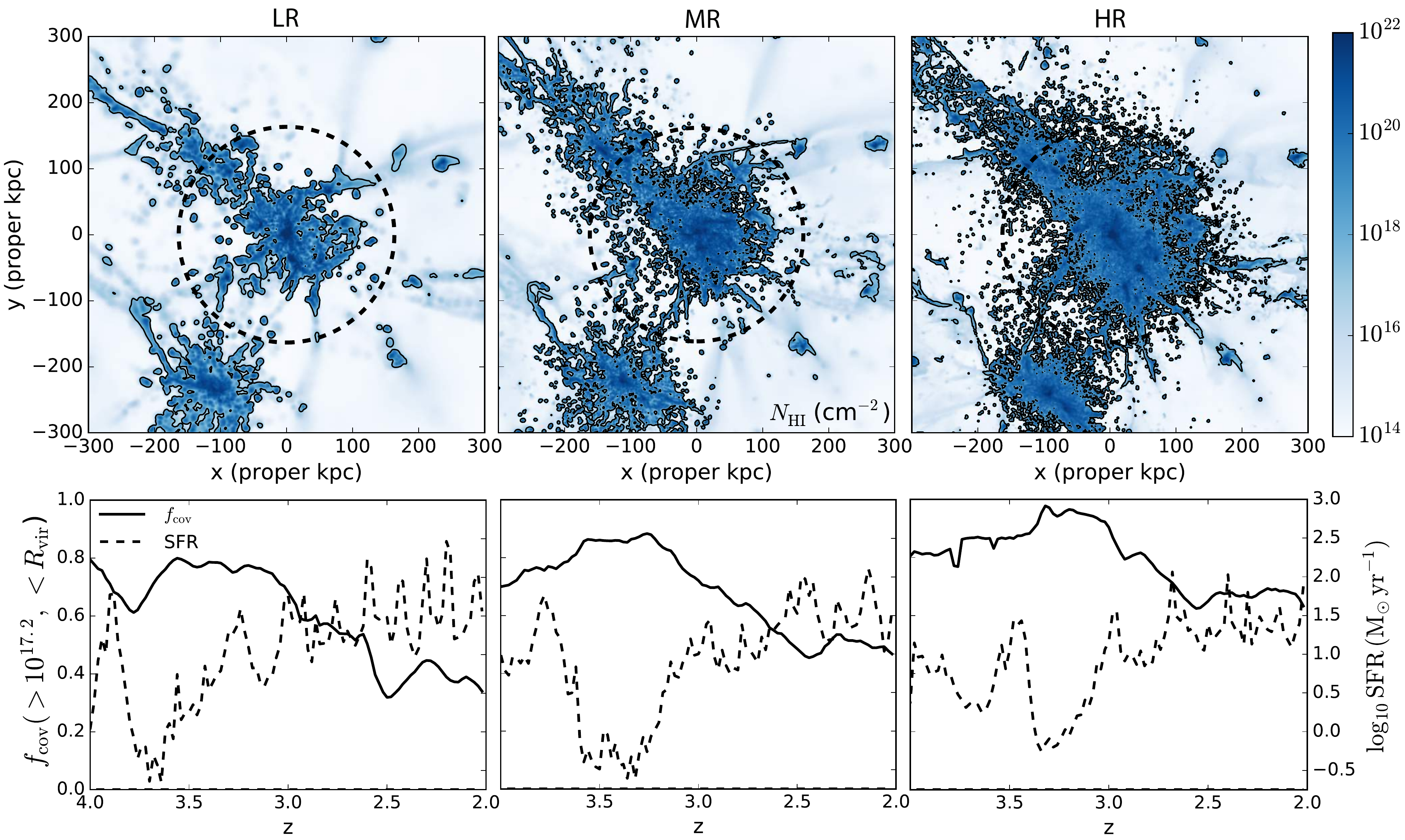}
\end{center}
\caption[]{\emph{Top:} $z=2$ HI maps for a $M_{\rm h}(z=2)=3.6\times10^{12}$ M$_{\odot}$ MassiveFIRE halo simulated at three resolution levels. \emph{Bottom:} LLS covering fraction and star formation rate within a virial radius from $z=4$ to $z=2$. The LLS covering fractions increase systematically with increasing resolution (from left to right), while the burstiness of the star formation history decreases with increasing resolution.}
\label{fig:conv_images_fcov} 
\end{figure*}

\vspace{-0.2in}
\section{Simulations and analysis methodology}
\label{sec:sims}
\subsection{Zoom-in simulations}
Our simulations implement the same stellar feedback physics and numerical methods as the ones analyzed in \citet{2014MNRAS.445..581H} and FG15; we refer to those papers for details. 
Briefly, the simulations were run using the GIZMO
simulation code  in P-SPH mode \citep[][]{2013MNRAS.428.2840H, 2015MNRAS.450...53H}. 
Gas is allowed to cool to $T\sim10$ K via atomic and molecular lines and star formation proceeds only in dense regions ($n_{\rm H}>5\,{\rm cm^{-3}}$) that are locally self-gravitating. 
Stellar feedback is modeled by implementing energy, momentum, mass, and metal return from radiation, supernovae, stellar winds, and photoionization following STARBURST99 \citep[][]{1999ApJS..123....3L}. 
During the course of the hydrodynamical calculation, ionization balance is computed using the ultraviolet background model of \cite{2009ApJ...703.1416F} and we apply an on-the-fly correction for self-shielded gas.

Our analysis in this paper combines the simulations previously analyzed in FG15 and new halos from the MassiveFIRE suite. 
The MassiveFIRE halos included in this analysis are the halos in the mass range $M_{\rm h} \approx 2\times10^{12}-10^{13}$ M$_{\odot}$ at $z=2$ introduced in Feldmann et al. (2016\nocite{2016MNRAS.458L..14F}). 
These halos span a wide range of accretion histories and environments. 
A subset of the MassiveFIRE simulations have been run at three resolution levels, labeled LR (low resolution), MR (medium resolution), and HR (high resolution). 
The HR simulations have a (zoom-in region) gas particle mass $m_{\rm b}=3.3\times10^{4}$ M$_{\odot}$ and a minimum (adaptive) gas gravitational softening $\epsilon_{\rm b}=9$ proper pc. The dark matter particle mass and gravitational softening lengths in the zoom-in regions are $m_{\rm dm}=1.7\times10^{5}$ M$_{\odot}$ and $\epsilon_{\rm dm}=143$ proper pc, respectively. 
These HR resolution parameters are similar to the `z2h' simulations and other LBG-mass halos analyzed in FG15. 
The MR simulations have the same gravitational softening parameters but higher gas and dark matter particle masses by a factor of 8 in zoom-in regions. 
The LR simulations have higher zoom-in particle masses by another factor of 8, double the minimum gravitational softening lengths of the MR and HR simulations, and a lower star formation density threshold of $n_{\rm H}=1$ cm$^{-3}$. 
Our final compilation of covering fractions is based on HR-level simulations only and we focus on $z=2-2.5$. 

In FG15, we concluded that simulations only including stellar feedback failed to explain the large LLS covering fractions observed around $z\sim2$ quasars. 
That conclusion was primarily based on our analysis of the m14 simulation ($M_{\rm h}(z=2) \approx 6\times10^{12}$ M$_{\odot}$).
The m14 simulation had a zoom-in gas particle mass $m_{\rm b}=4.4\times10^{6}$ M$_{\odot}$ much larger than the HR-level LBG-mass halos included in the analysis. 
We show in \S \ref{sec:resolution} that the m14 simulation did not have sufficient resolution to produce converged CGM predictions and so we exclude it from our updated analysis. 
We also exclude the m13 simulation analyzed in FG15 since its resolution was closer to MR level than HR level.

\subsection{CGM analysis methodology}
\label{CGM analysis method}
Our analysis is similar to that performed in FG15; we only summarize here the key points. 
To evaluate covering fractions, the particle data are first projected onto a Cartesian grid of side length $L$ centered on the halo with $N$ grid points along each dimension. 
In this paper, we focus on LLS covering fractions within a projected virial radius, defined as the fractions of projected pixels with HI column density $N_{\rm HI} > 10^{17.2}$ cm$^{-2}$.  
We use the \cite{1998ApJ...495...80B} virial radius definition. 
For the new massive halos, we use $L=600$ proper kpc and $N=600$, corresponding to a spatial grid resolution of 1 proper kpc. 
Our LLS covering fractions are well converged with grid resolution. 
To approximate neutral fractions in self-shielded gas, we use the analytic fits to radiative transfer calculations developed by \citet[][]{2013MNRAS.430.2427R}. 
We neglect ionization of CGM gas by local sources. 
This tends to overestimate HI covering fractions, but only slightly for LLSs around ordinary galaxies \citep[e.g.,][]{2011MNRAS.412L.118F, 2011MNRAS.418.1796F}. 
\cite{2007ApJ...655..735H} showed that the clustering of LLSs around luminous quasars is highly anisotropic, consistent with LLSs being photo-evaporated along the line of sight but largely unaffected by the quasar radiation in the transverse direction. 
For our comparison with LLSs transverse to quasars, we thus also neglect local ionization effects.

\vspace{-0.2in}
\section{The CGM of high-redshift massive halos}
\label{sec:results}

\subsection{Lyman limit system properties}
\label{sec:lls_properties}
Figure \ref{fig:halo_images} shows HI column density, gas-phase metallicity, and stellar surface density maps for three representative high-resolution halos from the MassiveFIRE sample. 
The halos are substantially filled with high-column and metal-enriched HI.  
The mean, median, and standard deviation of $\log_{10}{(Z/Z_{\odot})}$, where $Z$ is the HI-mass weighted metallicity, for LLS sight lines within a projected $R_{\rm vir}$ (but excluding the inner 20 proper kpc to minimize contamination from the central galaxy) are $-1.1$, $-0.9$, and 0.7, respectively (assuming $Z_{\odot}=0.14$; Asplund et al. 2009\nocite{2009ARA&A..47..481A}). 
The projected gas kinematics are complex (velocities up to $\sim 500$ km s$^{-1}$; see Fig. \ref{fig:halo_images}) and it is not generally possible to use LLS metallicity to cleanly separate cosmological inflows or galactic winds in an instantaneous sense (see also Hafen et al., in prep.). 
Overall, the metallicity and kinematic properties of dense HI in our simulated massive halos appear broadly consistent with observational constraints from high-dispersion spectra of the $z\sim2$ quasar CGM \citep[][]{2015arXiv151006018L}.  
Interestingly, the overall spatial distribution of LLSs correlates with the spatial distribution of satellite galaxies, indicating that satellites  play an important role in shaping the HI distribution in massive halos. 
As we showed for LGB-mass halos in FG15, ejection of cool gas by both central and satellite galaxies can interact with infalling large-scale structure filaments to enhance LLS covering fractions substantially. 

Figure \ref{fig:fcov_summary} summarizes the LLS covering fractions evaluated within a projected virial radius for the simulations previously analyzed in FG15 and for the new MassiveFIRE halos at $z=2$ and $z=2.5$.  
The simulated covering fractions are compared to the average covering fractions measured by \cite{2012ApJ...750...67R} around LBGs and by PHS13 in halos hosting quasars over matching redshift ranges. 
To facilitate the comparison of our simulated halos with the Rudie et al. and Prochaska et al. observational data points, we also show averages over the LBG- and quasar-mass halos in our simulation sample. 
The open black circle averages over all halos of mass $10^{11.5} \leq M_{\rm h} \leq 10^{12.5}$ M$_{\odot}$ ($f_{\rm cov}=0.26\pm0.11$) 
and the open black star averages over all MassiveFIRE halos ($M_{\rm h}\approx 10^{12}-10^{13}$ M$_{\odot}$; $f_{\rm cov}=0.55\pm0.13$).

Overall, we find good agreement between the simulated and observed covering fractions for both the LBG and quasar samples, with the covering fractions increasing systematically with increasing halo mass and from $z=2$ to $z=2.5$. 
We find much higher covering fractions than in FG15 in our more statistically robust and higher-resolution sample of quasar-mass halos. 
We explain this difference in the next section. 
Our predicted covering fractions in quasar-mass halos are also higher by a factor $\sim3$ than those predicted by the simulations of \cite{2014ApJ...780...74F}. 
Our simulated halos include strong stellar feedback but no AGN feedback, suggesting that the high covering fractions measured around quasars do not require a significant contribution from AGN feedback (although such feedback could certainly be important in real halos).

While the covering fractions appear to increase smoothly from the LBG mass regime to the quasar mass regime, the increase is rather steep: a factor $>2$ increase in covering fraction for factor of 5 increase in halo mass (computed from the averages shown by the open black symbols in Fig. \ref{fig:fcov_summary}). A few effects likely contribute to this steep increase. First, low-mass main halos have small HI covering fractions within their virial radii, as can be seen from the open blue symbols in Figure \ref{fig:fcov_summary}. When such low-mass halos are the satellites of more massive main halos, they do not contribute a large LLS cross section. Quasar-mass halos, however, can have satellites that are as massive as LBGs and which can each contribute a significant fraction of their virial cross section as LLSs. In addition, quasar hosting halos are massive enough that they may be able to sustain a quasi-static hot atmosphere \citep[][]{2003MNRAS.345..349B, 2005MNRAS.363....2K, 2011MNRAS.417.2982F} whose pressure may increase the density of cool gas and, more generally, change how cosmological inflows and outflows propagate inside halos.

In their analysis of the EAGLE simulations, \cite{2015MNRAS.452.2034R} were also able to reconcile the LLS covering fractions predicted by their simulations with those observed around quasars by PHS13. 
However, the agreement found by \cite{2015MNRAS.452.2034R} is due primarily to a different effect. 
In their simulations, the median LLS covering fraction within $R_{\rm vir}$ increases by less than 30\% from $M_{\rm h}=10^{12}$ M$_{\odot}$ to $M_{\rm h}=10^{13}$ M$_{\odot}$ at $z=2$. 
Instead, \cite{2015MNRAS.452.2034R} find agreement by assuming that all quasars in the PHS13 sample are hosted by halos with mass $M_{\rm h}>10^{12.5}$ M$_{\odot}$ and following the mass distribution of the most massive halos in the EAGLE box. 
They then compared PHS13's observations with simulated halos at fixed transverse proper projection, rather than as a fraction of $R_{\rm vir}$. 
Since LLS covering fractions decrease with increasing fraction of $R_{\rm vir}$, \cite{2015MNRAS.452.2034R} effectively obtained high covering fractions by assuming that many of PHS13's sight lines probe a smaller fraction of the virial radius of foreground quasars than assumed by PHS13, who reported covering fractions assuming a fixed virial radius appropriate for $M_{\rm h}\approx 10^{12.5}$ M$_{\odot}$. 
This highlights the fact that different state-of-the-art feedback models still predict significantly different HI covering fractions around massive high-redshift galaxies.

\subsection{Numerical convergence}
\label{sec:resolution}
In Figure \ref{fig:conv_images_fcov}, we compare HI maps for HR, MR, and LR runs for a representative $M_{\rm h}(z=2)=3.6\times10^{12}$ M$_{\odot}$ halo. 
The maps show that the LLS covering fractions increase systematically with increasing resolution. 
This is confirmed more quantitatively by the bottom panels, which show the corresponding covering fractions and star formation rates within the halo for 100 time slices between $z=4$ and $z=2$. 
An important factor determining the high resolution needed to obtain converged HI covering fractions is that it requires not only resolving the generation of galactic winds from central galaxies, but also from lower mass satellites that are represented by a smaller number of resolution elements.

The systematic increase in predicted LLS covering fractions with increased resolution is the most important factor driving the different conclusion that we reached previously (FG15) regarding quasar-mass halos. 
That analysis was based primarily on the covering fractions of the m14 simulation. 
Even the LR version of MF2 has slightly smaller gas particle mass and minimum gas softening length than m14 ($m_{\rm b}=2.1\times10^{6}$ vs. $m_{\rm b}=4.4\times10^{6}$, and $\epsilon_{\rm b}=18$ proper pc vs. $\epsilon_{\rm b}=70$ proper pc). 
The simulated LLS system covering fractions increase slightly even from the SR run to the HR run in Figure \ref{fig:conv_images_fcov}, indicating that covering fractions may not be fully converged even in our HR simulations for quasar-mass halos.  
We stress, however, that the majority of the LBG-mass halos analyzed in FG15 had resolution similar to the HR runs analyzed here and that FG15 demonstrated convergence of their HI covering fractions for those halos. 

Finally, it is worth noting that Figure \ref{fig:halo_images} shows that the dense HI distribution in our massive halos is clumpy. 
In detail, the phase structure of the CGM probably depends not only on the subgrid models for stellar feedback and resolution parameters, but also on the properties of the hydrodynamic solver employed \citep[e.g.,][]{2012MNRAS.425.2027K, 2013MNRAS.429.3341B} and whether ``non-ideal'' hydrodynamical effects such as magnetic forces and thermal conduction are included.  Such effects can, for example, significantly affect the survival of cool clouds in galactic winds \citep[e.g.,][]{2015MNRAS.449....2M}. 
It is thus prudent to regard the CGM phase structure (including the detailed temperature distribution of galactic winds) predicted by our simulations as uncertain. 
Nevertheless, our simulations provide a clear demonstration that an explicit implementation of stellar feedback processes that successfully explains the stellar masses of galaxies without any parameter tuning (Hopkins et al. 2014\nocite{2014MNRAS.445..581H}; Feldmann et al. 2016) also predicts the presence of sufficient cool gas in galaxy halos to explain LLS covering fractions around both LBGs and quasars at $z\sim2-2.5$. 

\vspace{-0.2in}
\section{DISCUSSION AND CONCLUSIONS}
\label{sec:conclusion}
Our central result is that the MassiveFIRE simulations, with strong stellar feedback but no AGN feedback, predict LLS covering fractions within a projected virial radius in good agreement with those measured by PHS13 around luminous quasars. 
In our simulations, the covering fractions are high in quasar-mass halos to large extent because stellar feedback drives galactic winds which interact with and expand cosmological filaments. 
It is thus critical for simulations to not only resolve the generation of galactic winds from central galaxies but also the winds from satellite galaxies embedded in associated large-scale structure. 

Our results suggest that AGN feedback is not necessary to explain the large covering fractions observed around quasars, though it is certainly possible that AGN feedback significantly affects the CGM of real quasars (e.g., Johnson et al. 2015\nocite{2015MNRAS.452.2553J}). 
One way to observationally test whether the presence of a luminous AGN affects the properties of halo gas on $\sim100$ proper kpc scales would be to obtain spectra transverse to foreground galaxies that inhabit halos of similar mass but do not have a luminous AGN.  
Such halos can be traced by highly star-forming sub-millimeter galaxies \citep[e.g.,][]{2012MNRAS.421..284H, 2015Natur.525..496N} or by $z\sim2$ galaxies selected based on their high stellar mass. 

\vspace{-0.2in}
\section*{Acknowledgments}
We grateful for useful discussions with Xavier Prochaska, Joe Hennawi, Ali Rahmati, Zach Hafen, Daniel Angl\'es-Alc\'azar, and Alexander Muratov. 
CAFG was supported by NSF grants AST-1412836 and AST-1517491, by NASA grant NNX15AB22G, and by STScI grants HST-AR-14293.001-A and HST-GO-14268.022-A.
RF was supported by NASA through Hubble Fellowship grant HF-51304.01-A. 
EQ was supported by NASA ATP grant 12-ATP-120183, a Simons Investigator award from the Simons Foundation, and the David and Lucile Packard Foundation. 
DK was supported by NSF grant AST-1412153. 
Support for PFH was provided by an Alfred P. Sloan Research Fellowship, NASA ATP grant NNX14AH35G, and NSF grants AST-1411920 and AST-1455342. 
The simulations analyzed in this paper were run on XSEDE computational resources (allocations TG-AST120025, TG-AST130039, and TG-AST140023) and on NASA High-End Computing resources (allocations SMD-14-5492, SMD-14-5189, and SMD-15-6530).

\vspace{-0.2in}
\bibliography{references}

\end{document}